\documentclass[9pt]{article}

\usepackage[utf8]{inputenc}
\usepackage{subfiles}

\usepackage{listings}
\usepackage{verbatim}
\usepackage{forest}
\usepackage[hidelinks]{hyperref}
\usepackage{multicol,caption}
\usepackage{geometry}
\usepackage{float}

\usepackage{graphicx}
\graphicspath{{./figures/}{../figures/}}

\title{Development of a NIC driver in C\#}
\author{Samuel Chassot\\EPFL, Switzerland\\
\href{mailto:samuel.chassot@epfl.ch}{samuel.chassot@epfl.ch}}
\date{}

\begin{document}


\maketitle

\begin{multicols}{2}

\section{Introduction}

Drivers have a special status among the developer community that sees them as mysterious and inaccessible. We think their extensive communication with the hardware and their need of high performance are the cause of this bad reputation. According to a widely held view, these two requirements cannot be met using high level languages. However high level languages' compilers and runtimes made great progress these past years to enhance the performance of programs. The use of these languages can also reduce by a significant amount the number of bugs and security issues introduced by the programmers by taking care of some error-prone parts like memory allocation and accesses. We also think that using high level languages can help to demystify the drivers' development.

With this project, we try to develop a driver for a network card, the Intel 82599, in C\#. Our goal is to find out the feasibility of such a development and the performance of such a driver. We will also be able to tell what could be missing today in C\# to write a driver. We base our driver on the model proposed by Pirelli (2020)\cite{tinynfsolal} and its implementation in C.

\section{Background}
Here is a background to understand how drivers and specially TinyNF work.
\subsection{Network Interface Controllers}
In this section, we present the overall architecture of Network Interface Controllers ("NICs") which is useful to understand the driver design and implementation.

\subsubsection{I/O}

The CPU and NIC use three main channels to communicate:
\begin{itemize}
    \item PCI registers: stored on the NIC, the CPU uses I/O ports to access them. They are used at start to configure the NIC.
    \item NIC registers: stored on the NIC too, the CPU accesses them through memory-mapped I/O.
    \item RAM: used to store packets and metadata, CPU accesses it as usual, the NIC accesses it through Direct Memory Access. They both need to poll the RAM to be aware of changes.
\end{itemize}

\subsubsection{Descriptors}

The CPU and the NIC use a datastructure stored in memory called a \textit{descriptor}. This structure contains (1) a pointer to a buffer in memory to store a packet and (2) some metadata including packet length and some flags. The pointer can be changed by the CPU so that it can manage packets pools if needed. The number of descriptors is fixed at initialization.
Descriptors are used to issue commands to the NIC: 
\begin{itemize}
    \item For the reception: the CPU gives a descriptor to the NIC indicating where to put the packet and the NIC gives it back after having stored the packet in the buffer and changed the metadata including the packet length.
    \item For the transmission: the process is similar, the CPU gives a descriptor to the NIC including the pointer to the packet data and metadata like packet length set by the CPU. The NIC gives it back once the packet is sent.
\end{itemize}

\subsubsection{Descriptors in Intel 82599}

\begin{figure}[H]
    \centering
    \includegraphics[width=\linewidth]{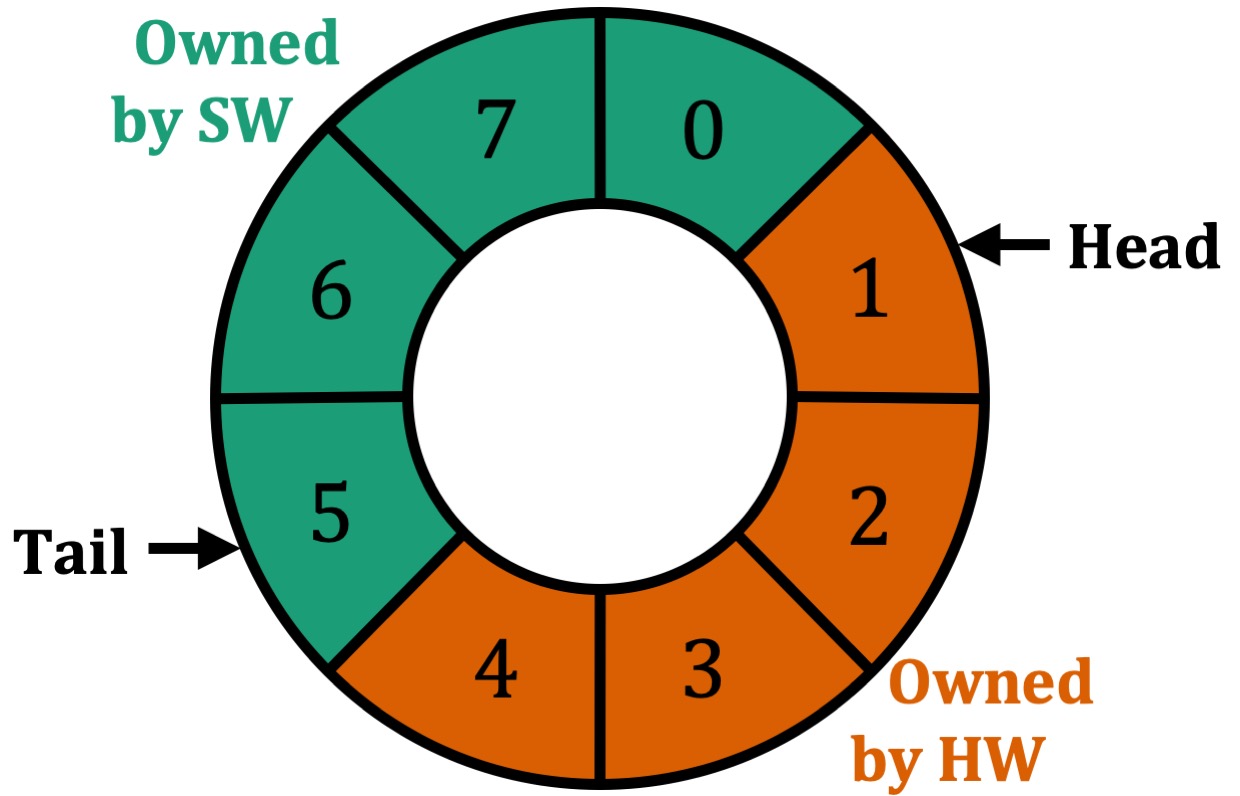}
    \captionof{figure}{A descriptor ring containing 8 elements. Descriptors 1 to 4 belong to the NIC while ones from 5 to 0 belong to the CPU.}
    \label{fig:single_ring}
\end{figure}

We present how descriptors are used by this specific card, for which our driver is developed.
As most of modern network cards, the Intel 82599 uses \textit{descriptors rings} to manage descriptors ownership between the CPU and the NIC (Figure \ref{fig:single_ring}). This ring is a region of memory (RAM) and two pointers, \textit{head} and \textit{tail}, that are located in NIC registers. The descriptors between the head inclusive and tail exclusive belong to the NIC while others belong to the CPU.
Each descriptor has a \textit{Done flag} in its metadata that the CPU used to differentiate uninitialized descriptors from those processed by the NIC.
To give a descriptor to the NIC, the CPU clears the \textit{Done flag} and increments the tail pointer. The NIC then sets the \textit{Done flag} and increments the head pointer to give it back.

Head and tail pointers can only be incremented. Decrementing them would be logically equivalent to steal descriptors.

\subsubsection{NIC queues}
A queue consists of a descriptor ring along with some configuration. This lets the NIC processing packets in parallel. By default all received packets go in the first receiving queue, developpers can then configure the NIC to put them in different queues based on their destination for example.

Transmission queues are just a way to have multiple CPUs handling packets from the same NIC without having synchronisation issues: a packet put in any transmission queue is sent in the same wire.

\subsection{TinyNF}

\subsubsection{Model}

We see in more details the model introduced by Pirelli (2020) \cite{tinynfsolal}: TinyNF.

This model is mainly purposed to be used with network functions (not in multi purposes computers). It could replace established ones like the one on which DPDK is based. These drivers are called \textit{kernel-bypass drivers} and they are based on the idea of using a fix amount of memory for buffers that is allocated at initialization time. This means that at no point during the execution a new buffer can be allocated. It completely removes the overhead of allocate memory during the execution (except at startup). The driver then manages a pool of free buffers which can be "allocated" (the memory is already allocated) to be used by the NIC to receive a new packet or by the network function to send a new packet. The function can also decide to keep some packets aside for some time to reconstruct a TCP message for example. It is allowed for the function to send new packet (without first having to receive one) too. This is a very flexible model. This flexibility comes with a cost though: complexity.

The main difference introduced by TinyNF to reduce this complexity is the way it deals with queues of descriptors (and so buffers). The idea is to remove the need of pool, which asks for management, by forcing a path for the buffer but also removing some possible actions of the function. 
We present one buffer cycle to illustrate. The buffer begins in the receiving queue, free. When a packet arrives, the NIC takes the first buffer of the queue, puts the packet in it, and moves the buffer in the processing queue. The processing unit (aka the network function) takes the first packet on the processing queue, apply its function to the packet and moves the buffer in the transmitting queue. In the transmission phase, the NIC takes the first buffer of the transmission queue and sends the packet through the wire and puts the buffer back in the receiving queue, ready to receive a new packet.

The transmission phase can be skipped by adding something in the metadata to indicate to the NIC not to send the packet and just to move the descriptor to the next queue (typically done by setting packet length to 0).
By using this "trick" the driver supports multi outputs. The driver has multiple transmission queues, one for each output, and each packet goes in all queues but the function chooses in which queue the packet is actually sent or not. These queues need only few synchronisation: the tails of all transmission queues are set at the same time and their heads are equal to the earliest of all queues' heads.

The model cannot support multiple receive queues as synchronising them is impossible but the whole system can be duplicated and run concurrently.
\begin{figure}[H]
    \centering
    \includegraphics[width=\linewidth]{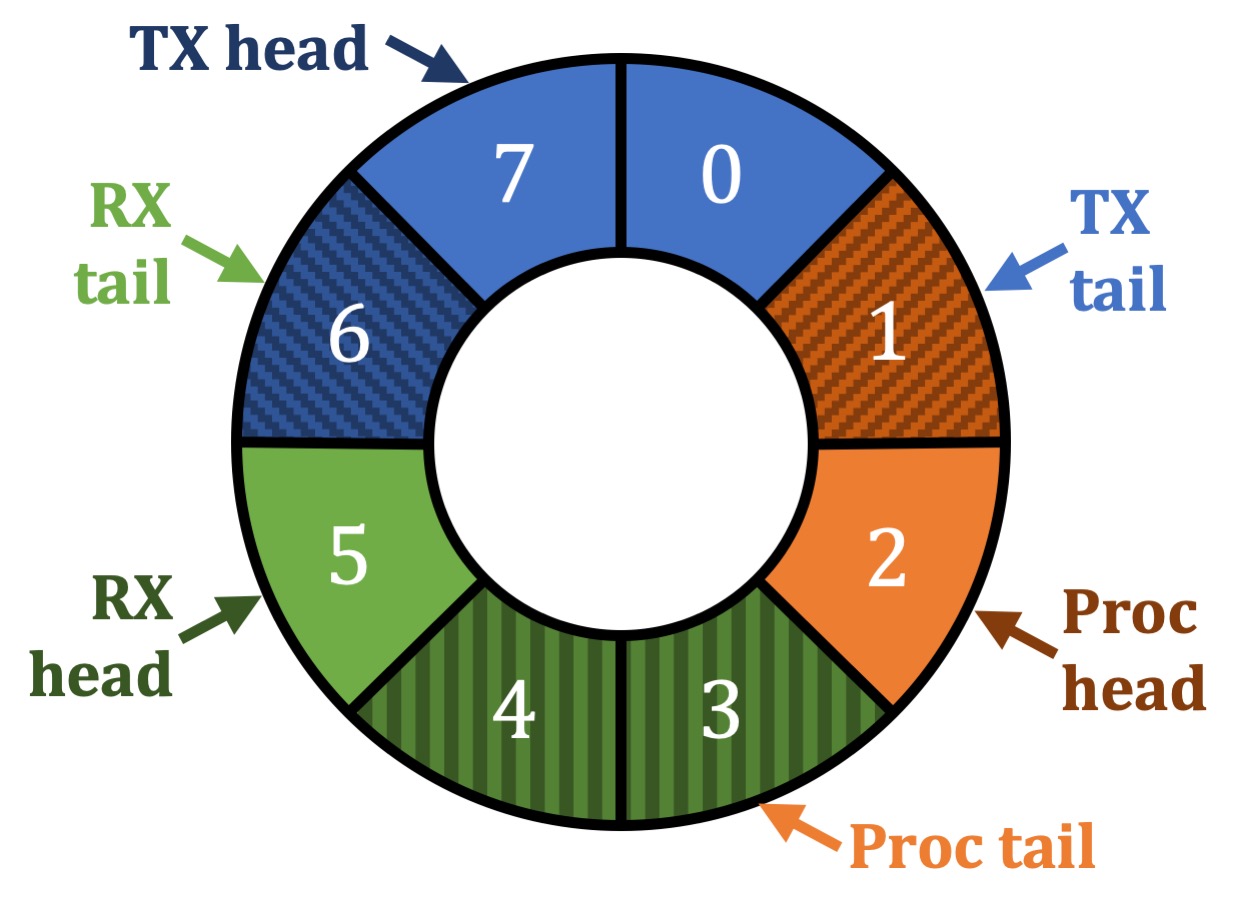}
    \captionof{figure}{A descriptor ring containing all queues. RX stands for receive, TX for transmission. Light ones are "in progress" while darker ones are "done". The head and tail pointers refer to "in progress" queues but implicitly define "done" queues.}
    \label{fig:virtual_ring}
\end{figure}
We see that this model reduces the flexibility of the driver: packets are processed in order, with no possibility of keeping some of them aside. This loss of functionality does not make this model useless: several of network functions that form the backbone of the internet (IP routing, Ethernet bridges for example) process packets one at a time without any reordering.
This has a great advantage in the complexity though, as it really brings down the number of execution path for a packet. Moreover, if we know that the network function always terminates, the driver cannot be stuck in a situation where all buffers are kept by the processing unit.

This also has a positive side effect: the model needs only one descriptors ring as the packets go through all steps in sequence with no reordering (Figure \ref{fig:virtual_ring}). At most one packet can be in the "processing" queue at any time.

\section{Implementation in C\#}
We use the .NETCore framework, an open source framework developed by Microsoft, available on Windows, Mac OS and Linux. The current version is .NetCore 3.1.3.
We now discuss interesting points encountered during the development.

Developing a driver in a high level language is seen as practically impossible or at least a bad idea. Drivers indeed need to communicate with the hardware and to have very high performance: two elements commonly thought as not achievable with high level languages. We try to show the opposite.
We unfortunately cannot do everything in C\# so we write a small C program that we use as an external library called directly in C\# code. Our goal is to minimize what is done in C to show what is missing today in C\# to fully develop a driver.

\subsection{Memory allocation}

A challenging issue when developing this kind of software in high level language is the memory allocation. Most of the time, the compiler takes care of it and it is beneficial for the developer. In this case, we need to allocate memory, in which the NIC can write and read, for the rings of descriptors and for the buffers containing the packets. This means that the memory needs to be managed by the developer. .NET provides a class to represent a \textit{MemoryMappedFile} \footnote{\url{https://docs.microsoft.com/fr-fr/dotnet/api/system.io.memorymappedfiles.memorymappedfile?view=netcore-3.1}} and it mimics the \textit{mmap} of C. The memory is allocated by creating a new \textit{MemoryMappedFile} that lives only in memory. We need to do a blank write to be sure the page is load in memory before trying to translate addresses. It is equivalent to adding the flag \textit{MAP\_POPULATE} when calling \textit{mmap} in C. We cannot call directly \textit{mmap} from \textit{stdio.h} because it has a parameter of type \textit{off\_t}. This type is not clearly defined and, as it does not exist in C\# and cannot be represented by a statically defined equivalent, we need to call a C function that calls \textit{mmap}. If we go see the code of the .NetCore framework runtime, we observe that it also uses a call to a C library to call \textit{mmap}. \footnote{\url{https://github.com/dotnet/runtime/blob/master/src/libraries/Common/src/Interop/Unix/System.Native/Interop.MMap.cs\#L33}}. This shows us that the .Net runtime also relies on C code, as we do in this project.  

To be sure that the required size is small enough to fit in one page, we need the value of \textit{\_SC\_PAGESIZE}. This value is defined as a macro in \textit{unistd.h} and its value can change from one machine to the other. We need to write a function in C that returns this value.

\subsection{Addresses translations}

As we have to communicate the physical address of the descriptors rings and buffers to the NIC so that it can access them, we need a way to translate virtual into physical addresses and the other way around. To achieve this, we need to open two files in the linux system, one for each way: \textit{/dev/mem} and \textit{/proc/self/pagemap} respectively. In C, it is done using \textit{mmap} so this should be doable using the same class \textit{MemoryMappedFile} in C\#. Unfortunately, with these special files, the call to \textit{MemoryMappedFile.CreateFromFile} fails. This is a known issue that should be addressed in the next release of .NETCore (see this issue\footnote{\url{https://github.com/dotnet/runtime/issues/26626}} and this one \footnote{\url{https://github.com/dotnet/runtime/issues/27638\#issue-370292487}}). So for now, this is done through a call to a function of the C library that opens the file and reads the wanted part.

\subsection{Communication with PCI devices}

As explained in the Background section, the driver needs to read and write registers that are located in the card itself, connected to the computer as a PCI device. To access it, the C code uses \textit{outl, outb, inl, inb} which are difined in \textit{sys/io.h}\footnote{\url{https://linux.die.net/man/2/outl}} as macros to x86 instructions used to interact with I/O ports. As they are macros, we cannot use them directly in C\# so we need to write a C function for each macro into the library.

\subsection{Volatile}

We use \textit{Volatile.Read} and \textit{Volatile.Write} as we translate the C code. Moreover, the memory model of the .Net framework is not very clear and its developers recommend using them for lock free accesses\footnote{\url{https://github.com/dotnet/runtime/issues/4906\#issuecomment-336464687}}.

\subsection{Memory accesses through pointers}

As the driver writes and reads memory at specific addresses, we have to use pointers. .Net introduced the \textit{Span\textless T\textgreater} class\footnote{\url{https://docs.microsoft.com/en-us/dotnet/api/system.span-1?view=netcore-3.1}} that represents a raw piece of memory beginning at a pointer and of a specific length, interpreted as type \textit{T} that could be used to read/write memory. We use pointer dereferencing instead because we use \textit{Volatile.Read} and \textit{Volatile.Write} and they work only with pointer dereferencing. To use \textit{Span\textless T\textgreater} instead, we tried to obtain a similar effect by adding a \textit{Thread.MemoryBarrier} before and after each read/write but it lowers the performance (around 16\% less throughput). With \textit{Span} but no barriers (no guarantees that the accesses won't be reordered by the compiler) it works but we have a small loss of throughput, so the risk taken by not putting barriers is not even worth it. As mentioned in the documentation\footnote{\url{https://docs.microsoft.com/en-us/dotnet/api/system.span-1.-ctor?view=netcore-3.1\#System_Span_1__ctor_System_Void__System_Int32_}}, the constructor performs no check but each access, read or write, does\footnote{\url{https://github.com/dotnet/runtime/blob/master/src/libraries/System.Private.CoreLib/src/System/Span.cs\#L142}}. 

As we want the best throughput, we use pointer dereferencing.

\subsection{Tiered Compilation}

The Tiered Compilation is a feature provided by the the JIT (Just-In-Time) compiler. The basic idea is to provide two versions of the same code: one version is lower-quality (aka less optimized),the second version is high quality (fully optimized). The first version is easier and so faster to produce for the JIT. The idea behind this is to start with the lower-quality version at startup and then replacing it if the method is used a given amount of time (around 30 calls) by the optimized version to save some JIT execution time. In this way, the time saved to generate the code dominates the loss of executing the less optimized version for a small amount of time. Then if the code is executed a lot, the savings of the optimized version become interesting and so the JIT replaces the code by the fully optimized one.\footnote{\url{https://github.com/dotnet/runtime/blob/master/docs/design/features/tiered-compilation.md}}

Quick JIT is related to Tiered Compilation: for pieces of code that do not contain loops and for which no pre-compiled parts exist, the JIT compiler produces code more quickly but with no optimisation at all. This saves time at startup but looses executing time\footnote{\url{https://docs.microsoft.com/en-us/dotnet/core/run-time-config/compilation\#quick-jit}}.

For a program like our driver, we are not interested in saving time at startup, we want the most optimized code. We are interested only at performance in the steady state so Tiered Compilation and Quick JIT are not relevant here. We disable both of them in this project.

As the benchmark performs a heat-up, enabling them or not should be the same but we observe a small loss of throughput (some 100s of Mbits/s) if we enable them. It does not change the performance if we change this setting for the code that is used only to setup the driver though. We disable them everywhere to be consistent.

\subsection{Weird bug at high speed}\label{ssse:weird_bug}
Passing a certain speed, the driver looses all packets and we do not exactly know why. By adding an annotation to disable compiler optimizations on the \textit{Receive} method the driver works as intended but the performance cannot be optimal.  We try to analyse the produced x86 assembly code, rewrite the C\# code differently, add memory barriers around critical memory operations but we cannot anything that could cause the problem. 

A suggestion comes from the developpers of \textit{Ixy}\footnote{\url{https://github.com/emmericp/ixy}}, an educational driver developed for the same card as ours, in the README of the repository on GitHub: 

"\textit{There's a weird problem on some systems that causes it to slow down if the CPU is too fast. DPDK had the same problem in the past. Try applying bidirectional traffic to the forwarder and/or underclock your CPU to speed up ixy.}"\footnote{\url{https://github.com/emmericp/ixy\#i-cant-get-line-rate-}} \\
We try to let the compiler optimizing the whole code and run the driver with some fixed CPU clock frequencies. We observe that the driver works normally up to 2.5Ghz but the bug appears at 2.7Ghz (as the CPU only accepts a set of frequencies, we cannot choose a frequency between 2.5Ghz and 2.7Ghz). This experiment shows us that the problem comes from the card itself and not from our code. As the throughput (see \ref{se:results}) is less than the version with the \textit{Receive} method not optimized but at maximum CPU clock frequency, we keep the latter.

\section{Results}\label{se:results}
We discuss the results of some benchmarks of our driver and compare its performance with some baselines'. The baselines are: the C version of TinyNF and DPDK ("unbatched").

The benchmark measures the following:
\begin{enumerate}
    \item the maximum throughput achieved with less than 0.1\% of lost packets.
    \item the latency for different throughputs, from 0 Mb/s to the maximum measured earlier, by increments of 1000 Mb/s.
\end{enumerate}
Unfortunately we cannot match the C version of TinyNF in performance, which obtains a maximum throughput of 20 Gb/s but our version achieves around 2/3 of that. We obtain nevertheless almost 2.5x more throughput than "unbatched" DPDK.

\begin{table}[H]
    \centering
    \begin{tabular}{| l | c | }
        \hline
         \multicolumn{2}{|c|}{\textbf{Maximum throughput in Mbits/sec}} \\
        \hline
        \hline
         \textbf{TinyNF C\#} & 12,695  \\ 
        \hline 
         \textbf{TinyNF C\#} &  \\ 
         \textbf{(full compiler optimisations, } & 12,148 \\
         \textbf{CPU at 2.5Ghz)} & \\
        \hline 
         \textbf{TinyNF C} & 20,000  \\  
         \hline 
         \textbf{DPDK no batched} & 5,781  \\  
         \hline
    \end{tabular}
    \caption{Maximum throughput with less than 0.1\% loss.}
    \label{tab:max_throughput}
\end{table}

Concerning the latency, we achieve very similar performance as TinyNF C version and "unbatched" DPDK. The C\# version follows the same "bump" around 11-12 Gb/s discussed by Pirelli (2020) in his paper \cite{tinynfsolal}.

\begin{figure}[H]
    \centering
    \includegraphics[width=\linewidth]{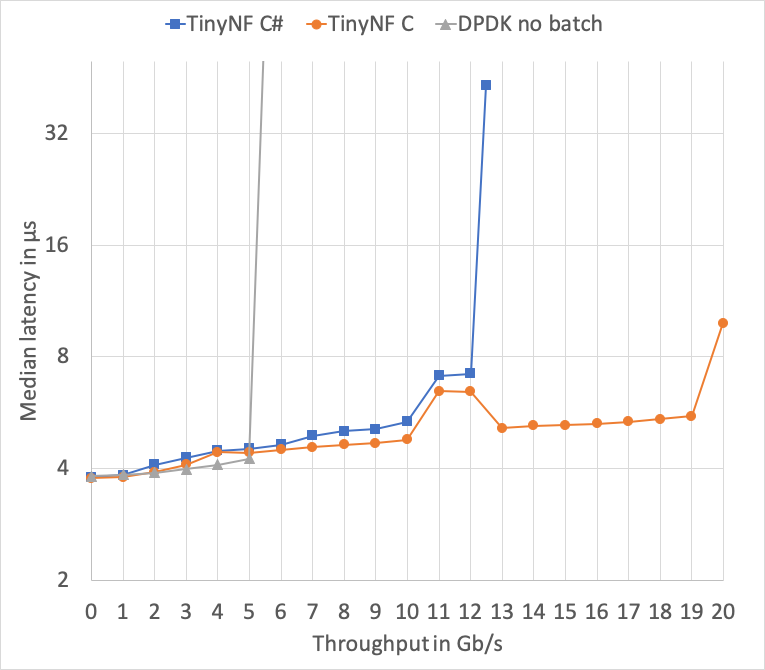}
    \captionof{figure}{Median latency for different throughput for the three drivers. The scale of the y axis is logarithmic.}
    \label{fig:latencies_median}
\end{figure}

\section{Conclusion}
The development of this driver in C\# shows several things:
\begin{enumerate}
    \item Developing a driver in C\# (a high level language) is possible with some very little calls to C for hardware accesses. This could even become possible without C calls in a near future.
    \item The performance achieved is significantly good.
\end{enumerate}

Majority of the community thinks that developing drivers in languages at higher level than C/C++ is, at least, a bad idea or simply impossible. With this project, we show that this is not the case. The compilers and runtime environments of high level languages as C\# made impressive progress these past years and manage to achieve really good performance while keeping development process less painful and less prone to errors than C/C++. We cannot achieve the same level of performance as a C version of the same driver but our implementation of a simpler model outperforms the DPDK standard implementation which is written in C. Moreover, by developing in high level languages, we can take advantage of the checks performed by the runtime on a part of memory accesses and of some features that are a lot simpler to write. A great example to illustrate the gain of productivity is how we pause the execution: in C, it requires a function of around twenty lines to wait x milliseconds; in C\# one line is sufficient by calling \textit{Thread.Sleep}. The environment of development of the new high level languages and their runtimes assist the developers in their task which leads to faster development process and less errors introduced in the program.

The performance could even be higher if we find a solution to the bug mentioned in \ref{ssse:weird_bug}. We know that the card is responsible for this bug but we do not know why it appears and why with the C\# version but not with the C version. For now, we force the compiler not to optimise a method, to overcome this bug because it gives better performance than underclocking.

Some parts need to be done in C for now. With the new version of .NetCore that will be released end of 2020, we will be able to remove two functions from the small C library: the two methods in charge of translating virtual-physical addresses and maybe, in the future, we will be able to remove entirely the C parts and do everything directly in C\#. For now, we think it is still worth the consideration.\\

Here is a link to the code: \url{https://github.com/samuelchassot/tinynf-csharp}

\end{multicols}
\newpage
\begin{multicols*}{2}

\end{multicols*}

\begin{thebibliography}{9}

\bibitem{tinynfsolal} 
Solal Pirelli and George Candea. 
\textit{A Simpler and Faster NIC Driver Model for Network Functions}. 
École Polythechnique Fédérale de Lausanne, 2020.
\end{thebibliography}
\end{document}